\documentclass[11pt,letterpaper]{article}

\usepackage{iftex}
\ifPDFTeX\usepackage[utf8]{inputenc}\fi
\usepackage[T1]{fontenc}
\usepackage[english]{babel}
\usepackage[left=0.76in,right=0.76in,top=0.68in,bottom=0.66in,footskip=0.35in]{geometry}
\usepackage{amsmath}
\usepackage{mathpazo}
\usepackage{microtype}
\usepackage{graphicx}
\usepackage{array,booktabs,caption}
\usepackage{xcolor}
\usepackage[numbers,sort&compress]{natbib}
\usepackage{placeins,float}
\usepackage{hyperref}
\usepackage{xurl}
\usepackage{doi}

\definecolor{paperlink}{HTML}{163477}
\hypersetup{colorlinks,linkcolor=paperlink,citecolor=paperlink,urlcolor=paperlink}
\newcolumntype{L}[1]{>{\raggedright\arraybackslash}p{#1}}
\newcolumntype{C}[1]{>{\centering\arraybackslash}p{#1}}
\newlength{\manuscripttablewidth}
\title{Subsurface Model Programming: Frontier Large Language Models Natively Interpret and Edit Dynamic Reservoir Models}
\author{Yanchen Song, Huizhu Xiang, Zhengdong Lei, Yuezhong Wang, Xingbang Liu}
\date{}
\renewenvironment{abstract}
  {\par{\centering\large\bfseries\abstractname\par}\small}
  {\par}
\hypersetup{pdftitle={Subsurface Model Programming: Frontier Large Language Models Natively Interpret and Edit Dynamic Reservoir Models},pdfauthor={Yanchen Song, Huizhu Xiang, Zhengdong Lei, Yuezhong Wang, Xingbang Liu}}

\makeatletter
\newcommand{\ClosingSectionSpacing}{%
  \setlength{\parskip}{4pt plus 1pt minus 1pt}%
  \renewcommand{\section}{\@startsection{section}{1}{\z@}%
    {-2.7ex \@plus -.8ex \@minus -.2ex}{1.5ex \@plus .2ex}{\normalfont\Large\bfseries}}%
  \renewcommand{\subsection}{\@startsection{subsection}{2}{\z@}%
    {-2.25ex \@plus -.7ex \@minus -.2ex}{1.2ex \@plus .2ex}{\normalfont\large\bfseries}}%
}
\makeatother

\begin{document}
\begingroup
\centering
{\fontsize{23}{27}\selectfont\bfseries Subsurface Model Programming:\par}
\vspace{4pt}
{\fontsize{17.5}{21}\selectfont\bfseries
Frontier Large Language Models Natively Interpret\\
and Edit Dynamic Reservoir Models\par}

\vspace{11pt}
{\large Yanchen Song\textsuperscript{*}, Huizhu Xiang, Zhengdong Lei,
Yuezhong Wang, Xingbang Liu\par}

\vspace{6pt}
{Research Institute of Petroleum Exploration \& Development\\
China National Petroleum Corporation, Beijing 100083, China\par}

\vspace{4pt}
{\small\url{https://github.com/syclong/llm-reservoir-engineering-benchmarks}\par}
\endgroup

\begingroup
\renewcommand{\thefootnote}{\fnsymbol{footnote}}
\footnotetext[1]{Corresponding author: Yanchen Song (\href{mailto:sycindex@gmail.com}{sycindex@gmail.com}).}
\endgroup

\vspace{11pt}
\begin{abstract}

\noindent
Multiphase flow simulation in heterogeneous porous media underpins decision-making across
hydrocarbon recovery, geological carbon sequestration, and geothermal extraction. These simulations
rely on input decks that specify grid topology, rock and fluid properties, and well controls. Yet
artificial intelligence has largely treated subsurface physics decks as black-box numerical
objects, relying on statistical surrogates or blind optimization across hundreds of forward runs.
Here we introduce Subsurface Model Programming, a workflow in which frontier reasoning large
language models (LLMs) read and modify raw simulation decks much as reservoir engineers inspect
source code. Without numerical gradients or external optimization algorithms, the LLM infers 3D flow
connectivity, energy depletion, and fluid breakthrough mechanisms directly from physical parameters
to propose targeted edits. Evaluated end-to-end in open-source multiphase simulators across four
benchmarks of increasing complexity, zero-tool full-file reasoning achieved 100\% accuracy in static
3D inter-well connectivity interpretation, correctly classifying all six reservoir-layer well-pair
labels and all three flow-path labels between open completions in a single pass. On PUNQ-S3, a single
proposal modified 54 of 1,761 active cells (3.1\%) and reduced well-balanced water-cut RMSE by 71.9\%.
On the SPE10 sector model, the best of five LLM proposals yielded 1.795 million $\text{sm}^3$ of
cumulative oil over ten years, outperforming a 25-evaluation particle swarm optimization baseline
by 8.9\%. Scaling to the 533,403-cell UNISIM-IV-2026 compositional model, five rounds of script-based
editing isolated dominant injector--producer channels, lowering combined field-and-well NRMSE from
37.98\% to 11.62\% (a 69.4\% reduction) and field production error by 73.9\%, with the final round's
improvement driven by a porosity edit along an existing injector--producer connection.
Because all model inputs, edits, and simulator outputs are archived, every result can be traced
directly to its underlying physical modification. Replacing exhaustive stochastic search with
physically auditable causal interventions in just one to five runs provides a generalizable
foundation for autonomous reservoir development, rapid history matching, legacy model diagnosis,
and coupled multiphysics engineering.

\end{abstract}

\nocite{Class2009,Pruess2004,Rasmussen2021,Zhong2021,Liu2026,Pan2024,Matejka2026,Samigullin2026,Onwunalu2010,Humphries2014,Naevdal2005,Oliver2011,Floris2001,Li2016ThiefZones,Botechia2022,Moyner2025,Clerc2002,Christie2001}

\section{Introduction}

Reservoir numerical simulation serves as the foundational computational engine for optimal hydrocarbon
recovery, geological carbon capture, utilization, and storage (CCUS) \citep{Class2009}, and deep
geothermal energy exploitation \citep{Pruess2004}, fundamentally solving coupled systems of non-linear
partial differential equations governing multiphase fluid flow across 3D heterogeneous geological grids
\citep{Rasmussen2021}. However, in the burgeoning landscape of Artificial Intelligence for Science (AI
for Science), current applications of artificial intelligence within subsurface engineering have
predominantly remained confined to peripheral, auxiliary workflows---such as training data-driven
surrogate or proxy models to emulate simulation outputs \citep{Zhong2021,Liu2026}, conducting
literature-based domain knowledge question-answering \citep{Pan2024}, or synthesizing scripts for
automated data pre- and post-processing \citep{Matejka2026,Samigullin2026}. Conversely, the raw
simulation decks that intrinsically encapsulate the full geological and physical reality---including 3D
corner-point grid geometries, spatial distributions of porosity and permeability tensors, fluid phase
equilibrium and PVT properties, and dynamic well production schedules---have long been treated as opaque
numerical ``black boxes'' inscrutable to large language models (LLMs). Lacking the capability to
natively interpret and edit these underlying physical files, core reservoir management tasks such as
history matching and field development planning remain heavily reliant on engineers' empirical
trial-and-error, or on physics-agnostic stochastic algorithms like Particle Swarm Optimization (PSO)
\citep{Onwunalu2010,Humphries2014} and Ensemble Kalman Filtering (EnKF) \citep{Naevdal2005,Oliver2011}.
These traditional methods necessitate hundreds or thousands of computationally prohibitive forward
simulations, lacking mechanistic intuition and frequently stagnating in suboptimal local extrema
\citep{Floris2001}.

This status quo gives rise to a foundational and frontier scientific question: Can frontier LLMs
transcend the superficial boundaries of conversational dialogue and auxiliary script generation to
natively interpret raw simulation decks as domain engineers and software developers read source
code---directly parsing 3D spatial coordinates, topological grid connectivity, and multiphase flow
mechanisms to autonomously execute high-fidelity parameter inversion and optimal development decisions?
In other words, without relying on external graphical pre/post-processing software or initiating massive
stochastic searches, can an LLM infer the in-situ dynamic flow state of subsurface fluids (e.g.,
discerning inter-well sandbody connectivity, identifying water-flooding thief zones
\citep{Li2016ThiefZones}, or diagnosing reservoir energy depletion) solely by ingesting raw model files,
and subsequently refactor the numerical simulation deck with programmatic precision?

In this study, through rigorous physical simulation benchmarks, we demonstrate for the first time that
frontier LLMs possess an emergent capability for Native Reading \& Editing of dynamic reservoir models,
propelling subsurface scientific computing from blind, black-box trial-and-error toward
physics-cognition-driven ``Subsurface Model Programming''. Centered on this breakthrough, we establish a
four-tiered hierarchical evaluation benchmark spanning static connectivity interpretation, single-step
history matching, ten-year field development optimization, and five-round closed-loop 3D compositional
property reconstruction. Crucially, all model-generated edits are coupled with high-performance
open-source reservoir numerical simulators for end-to-end physical verification. Specifically, our
primary findings and contributions are threefold:

\textbullet\ \textbf{First empirical demonstration of LLMs' native read-write capabilities on dynamic
reservoir models:} We demonstrate that frontier LLMs can directly manipulate dynamic reservoir models
analogous to writing and refactoring software code. For full simulation decks comprising tens of
thousands of grid cells, the model directly accomplishes 3D inter-well connectivity inference,
single-step parameter inversion, and field development planning via Deck-in-Context reasoning on the raw
deck files alone. For industrial-scale assets with millions of corner-point grid cells, the model
leverages lightweight diagnostic tools to generate structured patches that edit underlying property
files in situ. These findings unveil the embryo of ``Native Reservoir-Model Programming,'' demonstrating
reasoning efficiencies that substantially surpass human heuristic expectations and traditional
algorithms, offering transformative implications for subsurface optimization and physical
interpretability.

\textbullet\ \textbf{Establishment of a rigorous physical verification protocol via open-source
simulation engines:} Every deck edited or synthesized by the LLM is executed directly within open-source
reservoir numerical simulators, using genuine multi-phase production matching errors and ten-year
cumulative oil recovery as objective quantitative criteria. This protocol guarantees that all model
modifications strictly adhere to the governing partial differential equations of multiphase flow,
providing an uncompromised, physics-closed-loop benchmark of the model's true physical reasoning
capabilities.

\textbullet\ \textbf{Elucidation of mechanistic reasoning advantages to reshape subsurface engineering
workflows:} By dissecting the model's reasoning trajectories and iterative parameter adjustments, we
illuminate how LLMs leverage internalized reservoir physical priors---such as water breakthrough timing,
perforation production profiles, and thief-channel identification---to achieve targeted, monotonic
convergence with remarkably fewer simulation runs than conventional stochastic optimization algorithms.
This establishes a transformative methodological foundation for autonomous field development planning,
few-shot history matching, and white-box pathological auditing of legacy reservoir assets.

\section{Methodology and Experimental Setup}

\subsection{Reservoir Simulation Files, Concept Definition, and Interaction Modes}

A reservoir numerical simulation input file defines the physical and operational state of a subsurface
asset, including 3D grid geometry, rock porosity and permeability distributions, fluid properties,
initial equilibrium, and well operating schedules. While compact models reside in a single text file,
large 3D models split extensive grid coordinate and rock property arrays across multiple accompanying
data files.

Building on this representation, we define Subsurface Model Programming as a paradigm that treats
numerical reservoir simulation files as programmable computational objects. Under this paradigm,
subsurface flow behavior---such as 3D grid connectivity, fluid displacement, and pressure depletion---is
interpreted by directly reading raw model files, and dynamic production responses are altered by
directly editing rock property fields, well completions, and operating schedules, completely independent
of conventional numerical optimizers such as particle swarm optimization (PSO) or ensemble Kalman
filtering (EnKF). Here, the ``programmer'' is a general-purpose large language model (LLM) operating
directly from natural-language engineering instructions, without numerical gradients or adjoint
sensitivities.

\begin{figure}[H]
\centering
\includegraphics[width=\linewidth,height=312pt,keepaspectratio]{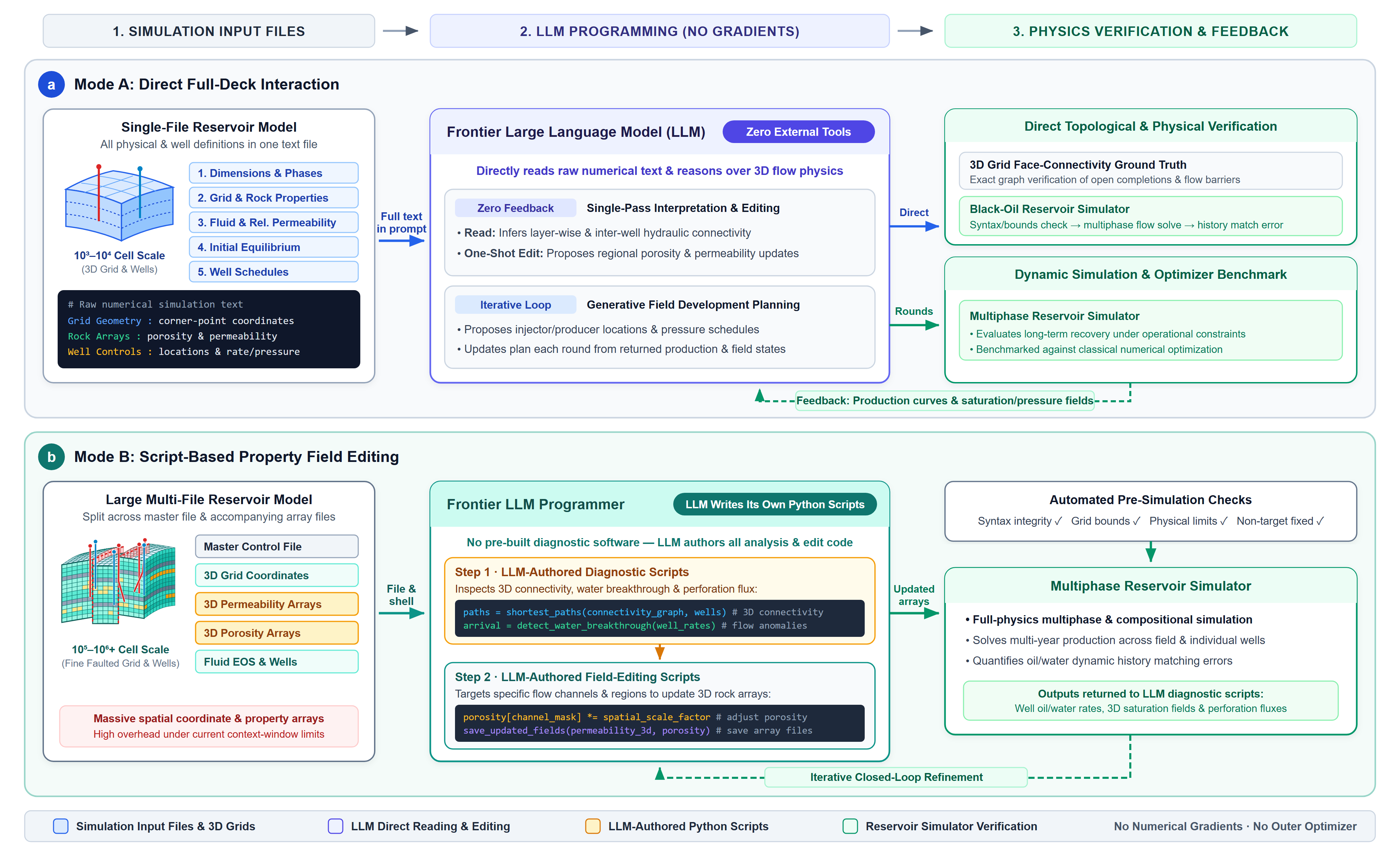}
\caption{Conceptual workflow of Subsurface Model Programming, illustrating direct full-deck interaction
(Mode A) and autonomous script-based property field editing (Mode B) coupled with reservoir numerical
simulators.}
\label{fig:framework}
\end{figure}

To systematically assess the capabilities and boundaries of large language models across tasks of
increasing physical complexity, we design four progressive benchmark experiments: static inter-well
connectivity interpretation (Exp 01), single-step black-oil history matching (Exp 02), multi-round
development planning (Exp 03), and multi-round compositional history matching (Exp 04). While an LLM
with a sufficiently long context window could theoretically ingest any reservoir model in raw text,
taking into account current context-length constraints, computational efficiency, and editing stability
across massive coordinate arrays, we implement these four tasks under two complementary interaction
modes (Fig.~\ref{fig:framework}):

\textbullet\ \textbf{Mode A --- Direct Full-Deck Interaction:} For compact single-file models comprising
thousands of grid cells (Exp 01--03), the entire simulation file containing all grid, rock, fluid,
initial-state, and well-schedule definitions is placed directly into the LLM prompt. In static
connectivity interpretation (Exp 01) and single-step history matching (Exp 02), the LLM operates with no
external tools, producing connectivity judgments or rock-property edits purely from text. In multi-round
development planning (Exp 03), the model updates well locations and pressure schedules over five rounds
using only simulator-returned production curves and saturation/pressure fields.

\textbullet\ \textbf{Mode B --- Script-Based Property Field Editing:} For large multi-file compositional
models---exemplified by UNISIM-IV \citep{Botechia2022}, which spans 19 files over a 47
\ensuremath{\times} 39 \ensuremath{\times} 291 corner-point grid totaling 533,403 grid cells (72,267
active cells) and millions of spatial coordinate and property entries---full-text ingestion remains
theoretically feasible given enough context, yet incurs substantial token overhead and reduced spatial
editing precision under current context limits. In this setting (Exp 04), using only a standard file and
command-line environment without pre-built reservoir diagnostic software, the LLM writes its own Python
scripts to analyze 3D grid connectivity, water-arrival timing, and perforation flux from simulator
outputs, and then writes targeted editing scripts to update 3D permeability and porosity files without
using any external optimizer or parameter sweep.

\clearpage

\subsection{Simulators, Models, and Evaluation Setup}

All model outputs are evaluated directly against multiphase flow physics using open-source reservoir
simulators. To match the keyword compatibility of different simulation models, complementary simulators
are used across tasks: OPM Flow (v2026.04) \citep{Rasmussen2021} runs black-oil history matching in Exp
02, whereas JutulDarcy \citep{Moyner2025} simulates ten-year development planning in Exp 03 and
five-round compositional history matching in Exp 04. For static interpretation in Exp 01, inferred
connectivity is verified directly against exact 3D geometric connectivity derived from adjacent
permeable active grid cells and flow barriers.

Without numerical gradients or external optimization algorithms, we select two reasoning models
according to task requirements: the flagship gpt-6-astra handles the zero-tool, full-deck tasks (Exp
01--03), while the more cost-efficient gpt-6.1-sol handles the multi-turn, script-based workflow (Exp
04) to test this capability across different model tiers. Every proposed edit passes automated checks
for keyword syntax, grid bounds, and physical parameter limits while keeping non-target settings fixed.
In Exp 03, LLM planning is compared under identical physical constraints \citep{Humphries2014} against
Particle Swarm Optimization (PSO) \citep{Clerc2002} at both 5- and 25-evaluation budgets.
Table~\ref{tab:benchmarks} summarizes the reservoir models, grid scales, tested LLMs, interaction modes,
and evaluation metrics across all four benchmarks.

\clearpage

\begin{table}[H]
\centering
\caption{Overview of the four reservoir simulation benchmarks.}
\label{tab:benchmarks}
\small
\renewcommand{\arraystretch}{1.16}
\setlength{\tabcolsep}{4pt}
\setlength{\manuscripttablewidth}{\dimexpr\linewidth-10\tabcolsep\relax}
\begin{tabular}{@{}L{0.122\manuscripttablewidth}L{0.202\manuscripttablewidth}L{0.130\manuscripttablewidth}L{0.155\manuscripttablewidth}L{0.176\manuscripttablewidth}L{0.215\manuscripttablewidth}@{}}
\toprule
\textbf{Experiment} & \textbf{Reservoir Model \& Grid Scale} & \textbf{Tested LLM} & \textbf{Interaction Mode} & \textbf{Task Objective} & \textbf{Simulator \& Evaluation Metric} \\
\midrule
\textbf{Exp 01} & Synthetic three-well anticline (19 \ensuremath{\times} 28 \ensuremath{\times} 3; 1,596 total / 1,146 active cells) & gpt-6-astra & Mode A: Direct Full\mbox{-}Deck (Read\mbox{-}Only, Zero Tools) & Infer layer-wise and inter-well connectivity & None (static geometric check); connectivity accuracy over adjacent permeable cells \\
\textbf{Exp 02} & PUNQ-S3~\citep{Floris2001} (19 \ensuremath{\times} 28 \ensuremath{\times} 5; 2,660 total / 1,761 active cells) & gpt-6-astra & Mode A: Direct Full\mbox{-}Deck (1\mbox{-}Round Edit, Zero Tools) & Single-step porosity and directional permeability update & OPM Flow 2026.04~\citep{Rasmussen2021}; field/well water-cut and oil/water rate errors \\
\textbf{Exp 03} & SPE10 sector model~\citep{Humphries2014,Christie2001} (60 \ensuremath{\times} 50 \ensuremath{\times} 1; 3,000 active cells) & gpt-6-astra & Mode A: Direct Full\mbox{-}Deck (5\mbox{-}Round Planning) & Optimize 4 well locations and 5 bottom-hole pressure stages over 10 years & JutulDarcy~\citep{Moyner2025}; 10-year cumulative oil production vs. 5- and 25-evaluation PSO~\citep{Clerc2002} \\
\textbf{Exp 04} & UNISIM-IV compositional model~\citep{Botechia2022} (47 \ensuremath{\times} 39 \ensuremath{\times} 291; 533,403 corner-point / 72,267 active cells across 19 files) & gpt-6.1-sol & Mode B: Script\mbox{-}Based Editing (5~Rounds) & Iterative 3D permeability and porosity field history matching & JutulDarcy~\citep{Moyner2025}; combined field and well normalized rate errors over 2,891 days \\
\bottomrule
\end{tabular}
\end{table}

\section{Experimental Results}

Evaluating large language models directly against multiphase flow simulators demonstrates that frontier
reasoning models can interpret raw reservoir simulation inputs, diagnose flow bottlenecks, and modify
three-dimensional rock properties with physical precision. Below, we present results across static
connectivity interpretation (Exp 01), single-step black-oil history matching (Exp 02), ten-year
development planning (Exp 03), and multi-round compositional field calibration (Exp 04).

\subsection{Zero-Tool 3D Inter-Well Connectivity Interpretation (Exp 01)}

\begin{figure}[H]
\centering
\includegraphics[width=\linewidth,height=396pt,keepaspectratio]{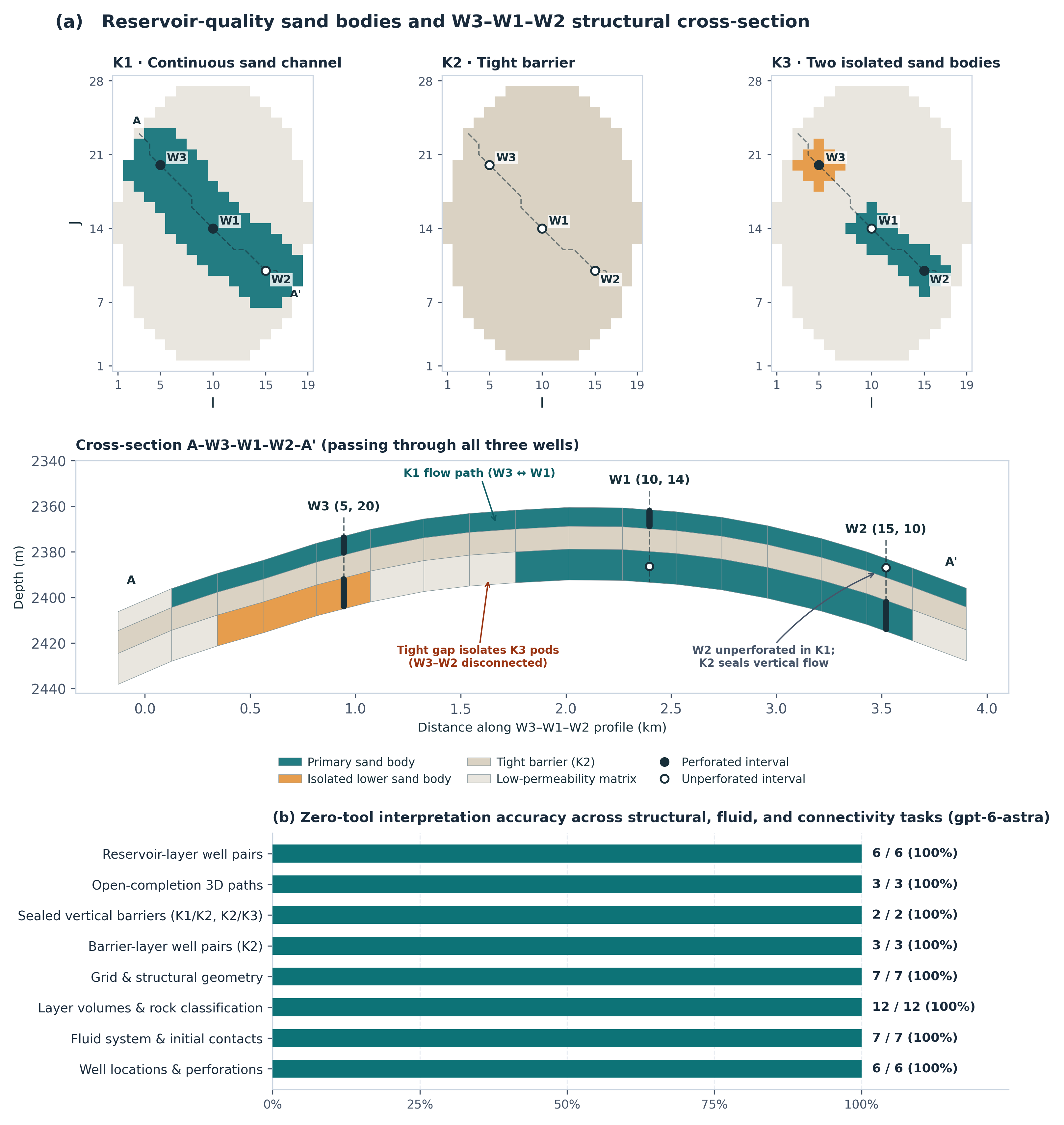}
\caption{Static 3D inter-well connectivity and zero-tool interpretation accuracy in Exp 01. (a)
Plan-view sand-body distributions across layers K1--K3 alongside the NW--SE structural cross-section
A--W3--W1--W2--A' passing through all three wells, showing continuous K1 connectivity between W3 and W1,
vertical sealing by K2, tight separation of the two K3 sand bodies, and open (filled) versus
unperforated (hollow) completions. (b) Item-level accuracy of gpt-6-astra across structural, fluid,
completion, and connectivity evaluation groups (100\% across all tasks).}
\label{fig:connectivity}
\end{figure}

Provided only with the raw single-file reservoir simulation input and no external visualization or
grid-parsing tools, gpt-6-astra achieved 100\% accuracy across all structural, fluid, completion, and
inter-well connectivity tasks (Fig.~\ref{fig:connectivity} and Table~\ref{tab:connectivity}). From
corner-point coordinates and tabular rock properties alone, the model accurately reconstructed the
three-layer anticlinal dome, exact grid dimensions, fluid contacts, and spatial geometry of all
high-permeability reservoir sand bodies.

\clearpage

Crucially, the model did not rely on naive visual or numerical shortcuts---such as assuming wells in the
same layer are automatically connected or treating trace background permeability as flow-conductive. As
shown in the W3--W1--W2 structural cross-section (Fig.~\ref{fig:connectivity}a), upper layer K1 forms a
continuous diagonal sand channel, middle layer K2 acts as a tight sealing barrier, and lower layer K3
splits into two separate sand bodies. The model correctly recognized that zero-transmissibility boundary
barriers isolate the sand bodies and seal the vertical interfaces across K2. Although well W3 is
perforated in both the upper and lower sand layers, crossflow inside the wellbore is disabled, and its
lower sand body is isolated from W2. The model therefore judged inter-well connectivity without error:
only W1 and W3 communicate through upper sand K1, whereas W2 (perforated solely in the isolated lower K3
body) remains hydraulically disconnected from both W1 and W3. All quantitative fluid contacts, well
positions, operating rate and pressure targets, and the 365-day schedule were likewise extracted without
error (Fig.~\ref{fig:connectivity}b).

\clearpage

\begin{table}[H]
\centering
\caption{Item-level verification summary for zero-tool static reservoir interpretation (Exp 01).}
\label{tab:connectivity}
\small
\renewcommand{\arraystretch}{1.16}
\setlength{\tabcolsep}{4pt}
\setlength{\manuscripttablewidth}{\dimexpr\linewidth-6\tabcolsep\relax}
\begin{tabular}{@{}L{0.244\manuscripttablewidth}C{0.122\manuscripttablewidth}C{0.123\manuscripttablewidth}L{0.511\manuscripttablewidth}@{}}
\toprule
\textbf{Verification Group} & \textbf{Correct / Total} & \textbf{Accuracy (\%)} & \textbf{Key Verified Findings} \\
\midrule
\textbf{Reservoir-layer well pairs (primary)} & 6 / 6 & 100.0 & Exact separation of continuous K1 channel and split K3 sand bodies \\
\textbf{Open-completion 3D paths} & 3 / 3 & 100.0 & W1--W3 connected via K1; W1--W2 \& W2--W3 disconnected \\
\textbf{Sealed vertical layer barriers} & 2 / 2 & 100.0 & Identified sealing barriers separating K1/K2 and K2/K3 interfaces \\
\textbf{Barrier-layer well pairs (K2)} & 3 / 3 & 100.0 & All well pairs across tight middle layer K2 labeled disconnected \\
\textbf{Grid \& structural geometry} & 7 / 7 & 100.0 & 19 \ensuremath{\times} 28 \ensuremath{\times} 3 grid (180 m spacing); central anticline crest beneath W1 \\
\textbf{Layer volumes \& rock types} & 12 / 12 & 100.0 & Exact reservoir vs. non-reservoir sand-body count and thickness per layer \\
\textbf{Fluid system \& contacts} & 7 / 7 & 100.0 & Black oil w/ dissolved gas; datum 2,355 m (234.46 bar); OWC 2,405 m \\
\textbf{Well locations \& perforations} & 6 / 6 & 100.0 & Exact well locations and perforated layer intervals for W1, W2, and W3 \\
\bottomrule
\end{tabular}
\end{table}

\subsection{Single-Step History Matching on PUNQ-S3 (Exp 02)}

\begin{figure}[H]
\centering
\includegraphics[width=\linewidth,height=504pt,keepaspectratio]{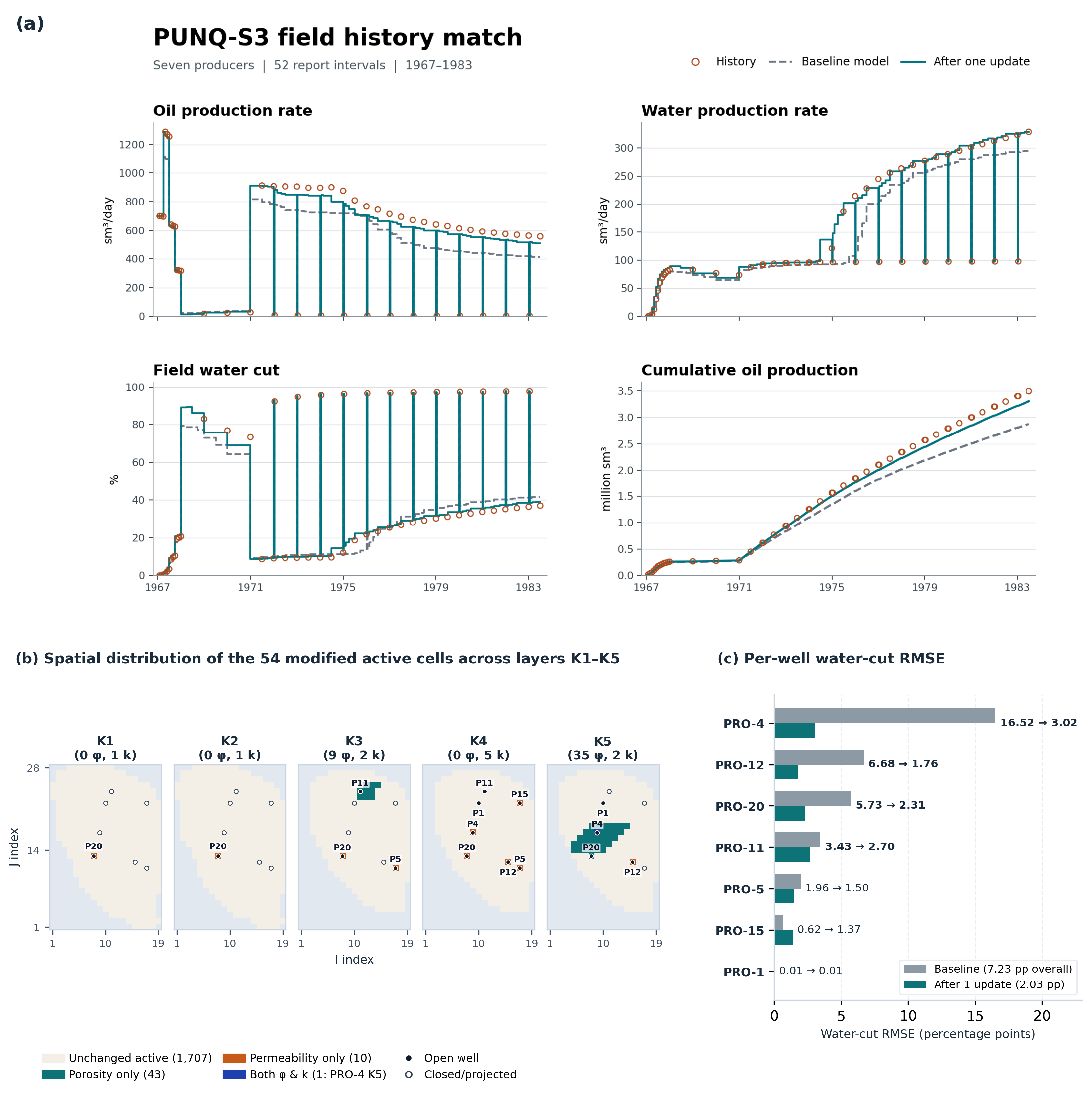}
\caption{Single-step history matching performance and spatial edit localization on PUNQ-S3 (Exp 02). (a)
Field-scale oil production, water production, water cut, and cumulative oil recovery comparing
historical observations, baseline, and the single-step candidate in OPM Flow (v2026.04). (b)
Layer-by-layer distribution (K1--K5) of the 54 modified active cells relative to producer locations. (c)
Per-well water-cut RMSE reduction across all seven producers.}
\label{fig:punq-edits}
\end{figure}

When asked to calibrate the PUNQ-S3 black-oil model in a single zero-tool inference pass, gpt-6-astra
proposed 16 localized property edits modifying just 54 of the 1,761 active grid cells (3.1\%)---updating
porosity in 44 cells and permeability in 11 cells, with one overlapping cell at the PRO-4 bottom
completion (Fig.~\ref{fig:punq-edits}b).

\begin{table}[H]
\centering
\caption{Field-scale and well-by-well history-matching error comparison before and after the single-step edit on PUNQ-S3 (Exp 02).}
\label{tab:punq-errors}
\small
\renewcommand{\arraystretch}{1.16}
\setlength{\tabcolsep}{4pt}
\setlength{\manuscripttablewidth}{\dimexpr\linewidth-8\tabcolsep\relax}
\begin{tabular}{@{}L{0.215\manuscripttablewidth}C{0.184\manuscripttablewidth}C{0.187\manuscripttablewidth}C{0.186\manuscripttablewidth}L{0.228\manuscripttablewidth}@{}}
\toprule
\textbf{Scope / Well} & \textbf{Water-Cut RMSE (pp) Before \ensuremath{\to} After} & \textbf{Oil Production RMSE (sm\textsuperscript{3}/d) Before \ensuremath{\to} After} & \textbf{Water Production RMSE (sm\textsuperscript{3}/d) Before \ensuremath{\to} After} & \textbf{Water-Cut Relative Change} \\
\midrule
\textbf{Well-balanced mean (primary)} & 7.2333 \ensuremath{\to} 2.0346 & 33.30 \ensuremath{\to} 12.10 & 8.78 \ensuremath{\to} 2.48 & \ensuremath{-}71.87\% \\
\textbf{Field total} & 3.8737 \ensuremath{\to} 1.2574 & 109.47 \ensuremath{\to} 33.65 & 26.81 \ensuremath{\to} 5.15 & \ensuremath{-}67.54\% (water prod. \ensuremath{-}80.80\%) \\
\textbf{PRO-4} & 16.5182 \ensuremath{\to} 3.0210 & 20.90 \ensuremath{\to} 3.82 & 20.90 \ensuremath{\to} 3.82 & \ensuremath{-}81.71\% \\
\textbf{PRO-12} & 6.6777 \ensuremath{\to} 1.7604 & 71.03 \ensuremath{\to} 31.12 & 5.81 \ensuremath{\to} 2.66 & \ensuremath{-}73.64\% \\
\textbf{PRO-20} & 5.7279 \ensuremath{\to} 2.3093 & 5.73 \ensuremath{\to} 2.31 & 5.73 \ensuremath{\to} 2.31 & \ensuremath{-}59.68\% \\
\textbf{PRO-11} & 3.4299 \ensuremath{\to} 2.7037 & 4.34 \ensuremath{\to} 3.42 & 4.34 \ensuremath{\to} 3.42 & \ensuremath{-}21.17\% \\
\textbf{PRO-5} & 1.9595 \ensuremath{\to} 1.4960 & 31.75 \ensuremath{\to} 1.89 & 4.17 \ensuremath{\to} 1.89 & \ensuremath{-}23.65\% \\
\textbf{PRO-15} & 0.6207 \ensuremath{\to} 1.3672 & 34.91 \ensuremath{\to} 4.55 & 0.48 \ensuremath{\to} 0.95 & +120.25\% (oil prod. RMSE \ensuremath{-}86.98\%) \\
\textbf{PRO-1} & 0.0105 \ensuremath{\to} 0.0142 & 0.013 \ensuremath{\to} 0.018 & 0.013 \ensuremath{\to} 0.018 & +35.44\% (+0.0037 pp) \\
\bottomrule
\end{tabular}
\par\smallskip
\begin{minipage}{\linewidth}
\footnotesize\raggedright \textit{Note:} Well-balanced water-cut RMSE averages over 274 flowing well/interval pairs across 7 producers; production RMSEs cover all 364 paired intervals. Field cumulative oil underestimation at year 16.5 fell from 17.77\% (2.87 million sm\textsuperscript{3}) to 5.47\% (3.30 million sm\textsuperscript{3} vs. 3.49 million sm\textsuperscript{3} observed).
\end{minipage}
\end{table}

\clearpage

Simulated in OPM Flow over 16.5 years of production history (1967--1983), this single-step adjustment
reduced every field-scale and balanced well-level history-matching error (Fig.~\ref{fig:punq-edits}a and
Table~\ref{tab:punq-errors}). The primary well-balanced water-cut error dropped by 71.9\% (RMSE
decreasing from 7.23 to 2.03 percentage points), while field-wide oil and water production errors fell
by 69.3\% and 80.8\%, respectively. As a result, simulated cumulative oil production rose closely toward
the historical benchmark (3.30 vs. 3.49 million sm\textsuperscript{3} observed), cutting cumulative
recovery underestimation from 17.8\% to 5.5\%.

\clearpage

\begin{figure}[H]
\centering
\includegraphics[width=\linewidth,height=474pt,keepaspectratio]{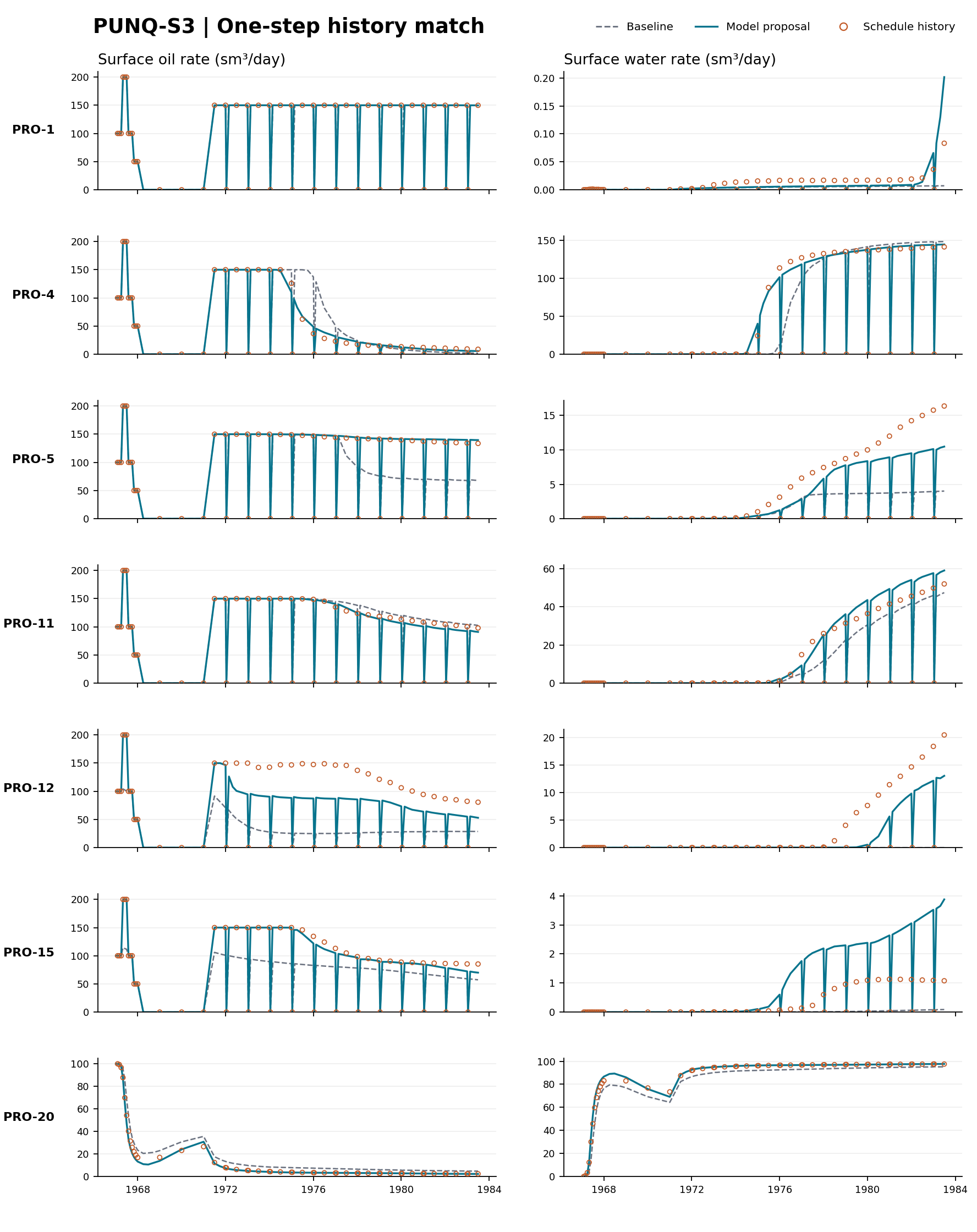}
\caption{Individual seven-producer surface oil and water production profiles before and after the
single-step LLM edit on PUNQ-S3 (Exp 02). Substantial oil production recovery occurs at PRO-5, PRO-12,
and PRO-15 alongside sharp water-breakthrough correction at PRO-4.}
\label{fig:punq-well-profiles}
\end{figure}

The spatial distribution of the 54 edited cells (Fig.~\ref{fig:punq-edits}b) and resulting well profiles
(Fig.~\ref{fig:punq-edits}c and Fig.~\ref{fig:punq-well-profiles}) reveal three clear, physically
grounded adjustments. First, to suppress premature bottom-water coning at PRO-4, the model lowered
porosity in a bottom-layer channel patch and reduced overly high permeability at the PRO-4 completion
while increasing horizontal permeability in the overlying K4 oil layer to sustain oil inflow, cutting
PRO-4 water-cut error by 81.7\% (RMSE from 16.52 to 3.02 percentage points). Second, to relieve low oil
production along the eastern and southern flanks, it boosted near-well horizontal permeability around
PRO-5, PRO-12, and PRO-15, sharply recovering oil production at all three wells. Third, localized
porosity reductions around PRO-11 and permeability reductions near the PRO-20 well interval curbed
excessive water production at both wells. Overall, five of seven producers improved in water-cut match;
only PRO-15 accepted slightly higher water cut in exchange for an 87.0\% reduction in oil-production
error, while nearly water-free PRO-1 showed negligible change (Table~\ref{tab:punq-errors}).

\clearpage

\subsection{Ten-Year Development Planning vs. PSO (Exp 03)}

\begin{figure}[H]
\centering
\includegraphics[width=\linewidth,height=447pt,keepaspectratio]{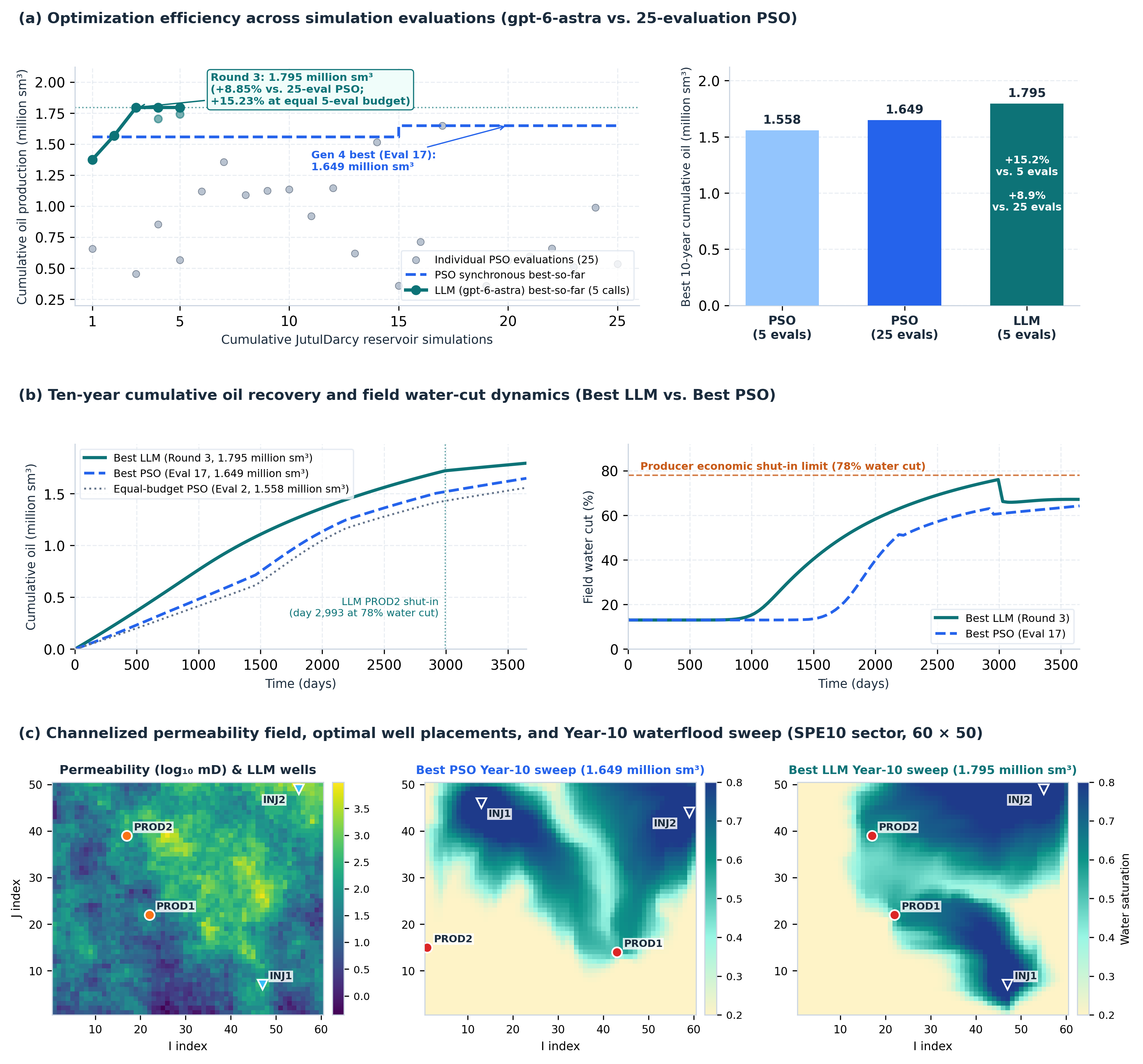}
\caption{Ten-year field development optimization performance on the channelized SPE10 benchmark (Exp
03). (a) Cumulative oil recovery trajectories across simulation evaluations, showing gpt-6-astra
surpassing both 5- and 25-evaluation PSO baselines by Round 3. (b) Ten-year cumulative oil production
and field water-cut history comparing Best LLM (Round 3) and Best PSO (Eval 17). (c) Log-permeability
channel architecture and Year-10 water saturation sweep maps for Best PSO versus Best LLM.}
\label{fig:spe10-planning}
\end{figure}

On the strongly heterogeneous SPE10 sector benchmark---jointly optimizing 28 decision variables (the
planar locations of two injectors and two producers plus five 2-year bottom-hole pressure stages per
well over a 10-year horizon under a 78\% economic water-cut shut-in limit)---gpt-6-astra outperformed
synchronous Particle Swarm Optimization (PSO) within only five reservoir simulations
(Fig.~\ref{fig:spe10-planning}a and Table~\ref{tab:spe10-planning}). Its best plan (Round 3) achieved a
ten-year cumulative oil production of 1.795 million sm\textsuperscript{3}, surpassing both the equal
5-simulation budget PSO best (1.558 million sm\textsuperscript{3}, +15.2\%) and the full 25-simulation
budget PSO search (1.649 million sm\textsuperscript{3}, +8.9\%) using five times fewer simulations.

\clearpage

Inspecting the well placements and final saturation fields explains why direct physical reasoning
outperformed population-based search (Fig.~\ref{fig:spe10-planning}b--c). In Round 1, placing wells too
closely along a high-permeability streak caused rapid water channeling and forced both producers to shut
in within two years (1.375 million sm\textsuperscript{3}). Guided by simulator feedback, the LLM
separated the two injectors into opposite high-permeability eastern corners at maximum 450 bar injection
pressure and positioned both producers inside the central-western high-permeability channels at minimum
100 bar bottom-hole pressure---maximizing pressure drawdown and driving broad east-to-west diagonal
waterflood sweep (Fig.~\ref{fig:spe10-planning}c, right). By contrast, PSO's best solution stranded
PROD2 near the low-permeability western edge (contributing only \textasciitilde{}10\% of total oil) and
left substantial bypassed oil at lower average reservoir pressure (Fig.~\ref{fig:spe10-planning}c,
center).

\begin{table}[H]
\centering
\caption{Summary of 10-year development planning proposals from gpt-6-astra (5 simulations) versus PSO (25 simulations) in JutulDarcy (Exp 03).}
\label{tab:spe10-planning}
\small
\renewcommand{\arraystretch}{1.16}
\setlength{\tabcolsep}{4pt}
\setlength{\manuscripttablewidth}{\dimexpr\linewidth-10\tabcolsep\relax}
\begin{tabular}{@{}L{0.141\manuscripttablewidth}C{0.125\manuscripttablewidth}C{0.130\manuscripttablewidth}C{0.095\manuscripttablewidth}L{0.254\manuscripttablewidth}L{0.255\manuscripttablewidth}@{}}
\toprule
\textbf{Method / Round} & \textbf{Simulation Budget / Evals} & \textbf{10-Year Cumulative Oil (million sm\textsuperscript{3})} & \textbf{Relative to Best LLM} & \textbf{Well Layout \& Displacement Strategy} & \textbf{Key Flow Behavior \& Shut-In Timing} \\
\midrule
\textbf{LLM Round 1} & 1 & 1.375 & \ensuremath{-}23.37\% & Clustered near fast channel & Rapid breakthrough; shut-in at days 328.5 \& 693.5 \\
\textbf{LLM Round 2} & 2 & 1.569 & \ensuremath{-}12.58\% & Injectors separated east; producers moved west & Eliminated early shut-in; surpassed 5-eval PSO best \\
\textbf{LLM Round 3 (Best)} & \textbf{3} & 1.795 & Best & East-corner injectors + central-western channels (450/100 bar) & Broad diagonal sweep; PROD2 shut-in late at day 2,993 \\
\textbf{LLM Round 4} & 4 & 1.704 & \ensuremath{-}5.04\% & Shifted PROD1 northeast toward central ridge & Earlier water breakthrough shut both producers in \\
\textbf{LLM Round 5} & 5 & 1.740 & \ensuremath{-}3.06\% & Returned to R3 layout with staged PROD2 pressure ramp & Delayed PROD2 shut-in until day 3,613.5 \\
\textbf{PSO Best (5-Eval Budget, Gen 1)} & 5 & 1.558 & \ensuremath{-}13.21\% & Northern injectors + south/west-edge producers & Best within 5-eval budget (Gen 1 mean: 0.818 million sm\textsuperscript{3}) \\
\textbf{PSO Best (25-Eval Budget, Gen 4)} & 25 & 1.649 & \ensuremath{-}8.13\% & Northern injectors + eastern producer; PROD2 at tight west edge & PROD2 stranded (\textasciitilde{}10\% of total oil); unswept flanks \\
\bottomrule
\end{tabular}
\end{table}

\subsection{Five-Round Compositional Reservoir History Matching (Exp 04)}

\begin{figure}[H]
\centering
\includegraphics[width=\linewidth,height=462pt,keepaspectratio]{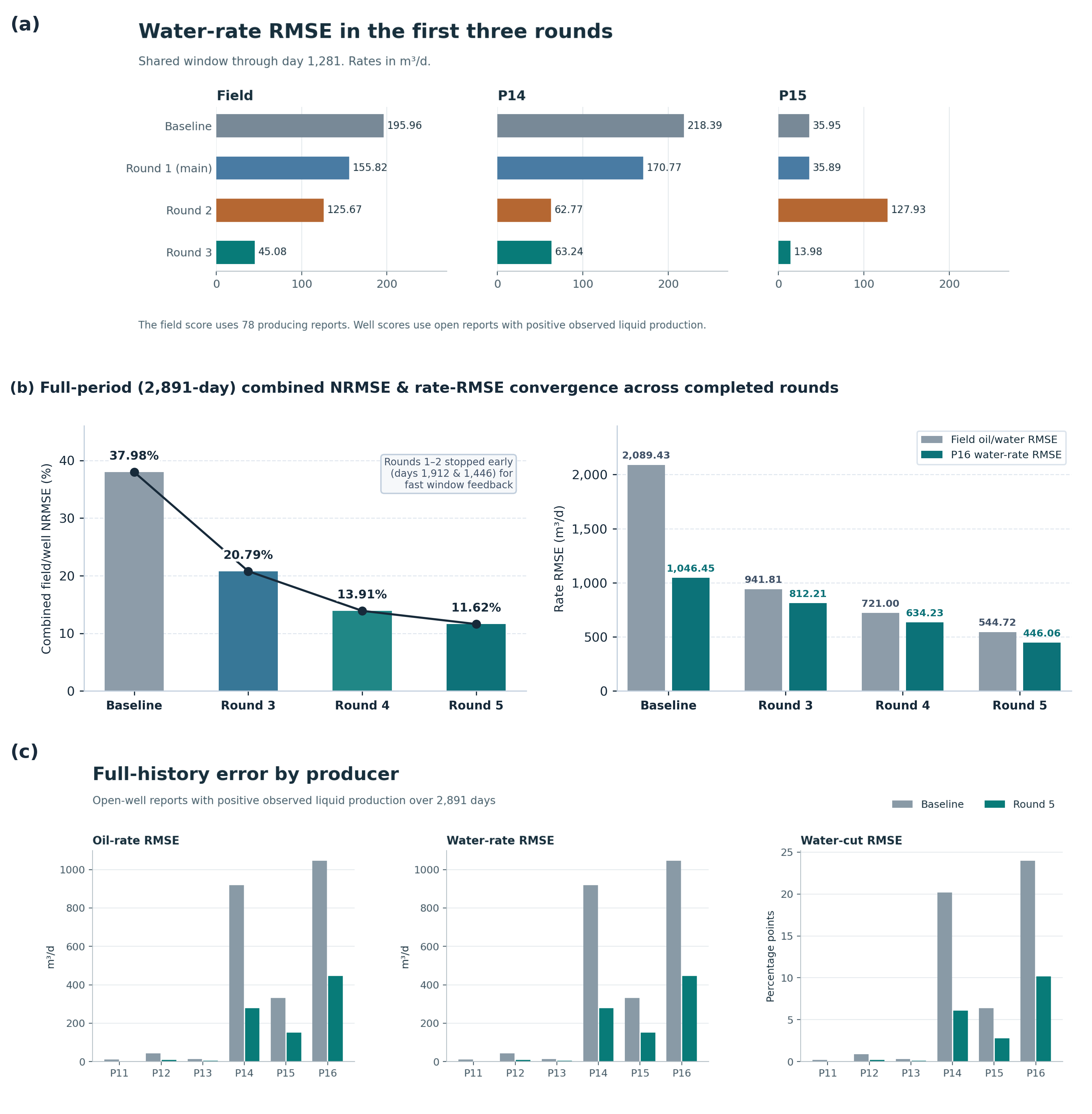}
\caption{Multi-round error convergence in the UNISIM-IV compositional history-matching benchmark (Exp
04). (a) Water production RMSE over the shared 1,281-day window across baseline and Rounds 1--3, showing
rapid P14 improvement and correction of Round 2's premature water at P15. (b) Monotonic full-period
(2,891-day) reduction in combined field/well NRMSE (37.98\% to 11.62\%) and field / P16 production RMSE
across completed rounds. (c) Full-history per-producer error reduction from Baseline to Round 5.}
\label{fig:unisim-convergence}
\end{figure}

On the 533,403-cell (72,267 active-cell) UNISIM-IV compositional benchmark, gpt-6.1-sol wrote and
executed its own diagnostic scripts over five iterative rounds to inspect three-dimensional flow
pathways and water-arrival timing before modifying rock property fields. Across the 2,891-day
(\textasciitilde{}7.9-year) production history, combined field-and-well normalized error (NRMSE) fell
monotonically across every completed round---from 38.0\% at baseline to 20.8\% in Round 3, 13.9\% in
Round 4, and 11.6\% in Round 5, an overall relative reduction of 69.4\%---while field-level oil and
water production error dropped by 73.9\% (2,089 to 545 m\textsuperscript{3}/d;
Fig.~\ref{fig:unisim-convergence}b and Table~\ref{tab:unisim-rounds}).

\clearpage

\begin{figure}[H]
\centering
\includegraphics[width=\linewidth,height=518pt,keepaspectratio]{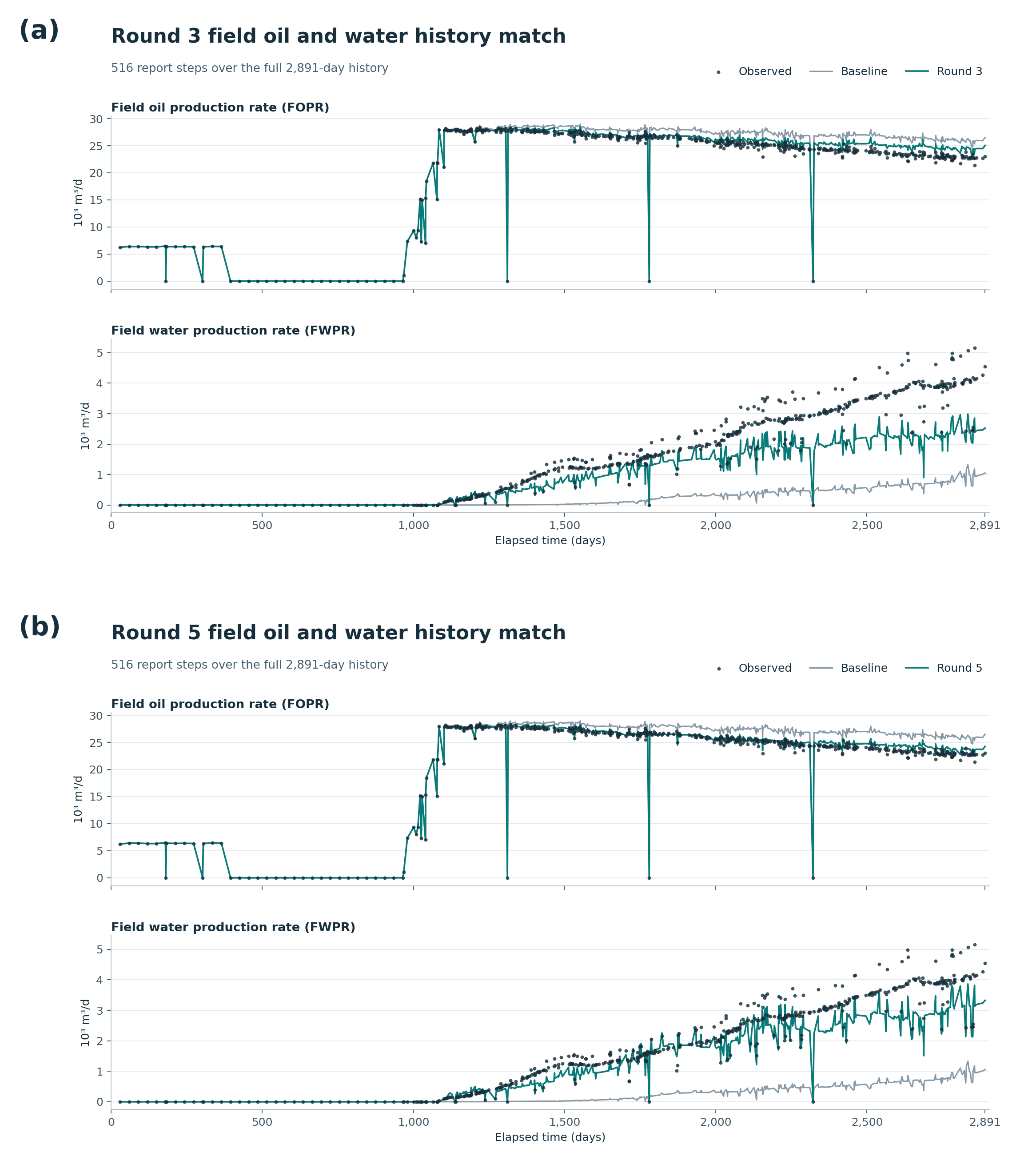}
\caption{Full-period (2,891-day, 516-step) field oil and water production history match in Exp 04
comparing (a) Round 3 (combined NRMSE 20.79\%) and (b) Round 5 (combined NRMSE 11.62\%) against observed
history and the baseline model.}
\label{fig:unisim-field-match}
\end{figure}

Rather than tuning parameters blindly, the five rounds exhibit a structured, self-correcting diagnostic
loop (Fig.~\ref{fig:unisim-convergence}a, Fig.~\ref{fig:unisim-field-match}, and
Table~\ref{tab:unisim-rounds}). During Rounds 1 and 2, partial simulations up to mid-history already
revealed clear diagnostic trends---showing that an exploratory shift in initial hydrocarbon composition
was unnecessary and that aggressive lower-zone permeability boosts caused premature water breakthrough
at P15---allowing both runs to stop early to save compute before scaling back those edits in Round 3 to
complete the full 2,891-day history at 20.8\% combined NRMSE. When P16's mid-to-late water surge after
neighboring injector I12-W startup reversed Round 3's error from slight early overprediction to severe
late underprediction, Round 4 replaced earlier bottom-water edits with enhanced conductivity along the
discrete I12-W--P16 channel (13.9\% NRMSE). Round 5 then held permeability fixed and lowered porosity
along that injector--producer pathway to accelerate flood-front advance
(Fig.~\ref{fig:unisim-spatial-edits}b), doubling P16's breakthrough response and reaching 11.6\% overall
NRMSE.

\clearpage

\begin{table}[H]
\centering
\caption{Multi-round progression of candidate modifications, simulation coverage, and full-period history-matching errors in UNISIM-IV (Exp 04).}
\label{tab:unisim-rounds}
\small
\renewcommand{\arraystretch}{1.16}
\setlength{\tabcolsep}{4pt}
\setlength{\manuscripttablewidth}{\dimexpr\linewidth-10\tabcolsep\relax}
\begin{tabular}{@{}L{0.087\manuscripttablewidth}L{0.245\manuscripttablewidth}C{0.122\manuscripttablewidth}C{0.158\manuscripttablewidth}C{0.108\manuscripttablewidth}L{0.280\manuscripttablewidth}@{}}
\toprule
\textbf{Round} & \textbf{Physical Modification Strategy} & \textbf{Simulated Window} & \textbf{Field Oil/Water Production RMSE (m\textsuperscript{3}/d)} & \textbf{Combined NRMSE} & \textbf{Diagnostic Outcome} \\
\midrule
\textbf{Baseline} & None (initial reservoir model) & 2,891 d (516 steps) & 2,089.43 & 37.9776\% & Delayed breakthrough at P14--P16; field water NRMSE 51.23\% \\
\textbf{Round 1} & Near-well permeability, bottom-water paths \& exploratory composition test & Stopped day 1,912\textsuperscript{*} & --- (1,281-d: 155.82 vs. 195.96) & --- & Early P14 improvement (218.39 \ensuremath{\to} 170.77); showed composition shift unnecessary \\
\textbf{Round 2} & Inherited R1 permeability with restored normalized composition; strong P14--P16 lower-zone boosts & Stopped day 1,446\textsuperscript{*} & --- (1,281-d: 125.67 vs. 195.96) & --- & P14 error dropped to 62.77, but triggered premature water at P15 (127.93) \& P16 \\
\textbf{Round 3} & Scaled back excessive P15/P16 lower-zone permeability \& porosity edits & 2,891 d (516 steps) & 941.81 & 20.7914\% & Eliminated premature P15 water (13.98); late P16 water still underpredicted \\
\textbf{Round 4} & Reverted P16 bottom-water edit; enhanced discrete I12-W--P16 channel & 2,891 d (516 steps) & 721.00 & 13.9150\% & Captured injector--producer sweep pathway; P16 full-period RMSE 634.23 \\
\textbf{Round 5} & Fixed permeability; reduced porosity only along I12-W--P16 path (1,120 cells) & 2,891 d (516 steps) & 544.72 & 11.6202\% & Best model (\ensuremath{-}69.40\% NRMSE); accelerated flood front to P16 (RMSE 446.06) \\
\bottomrule
\end{tabular}
\par\smallskip
\begin{minipage}{\linewidth}
\footnotesize\raggedright \textsuperscript{*}Rounds 1 and 2 were halted cleanly without solver errors once partial-window diagnostics clarified physical trends; field oil and water production RMSEs in m\textsuperscript{3}/d are equal under historical liquid-rate control.
\end{minipage}
\end{table}

\begin{table}[H]
\centering
\caption{Field-scale and producer-level full-period (2,891-day) history-matching performance in UNISIM-IV (Exp 04).}
\label{tab:unisim-errors}
\small
\renewcommand{\arraystretch}{1.16}
\setlength{\tabcolsep}{4pt}
\setlength{\manuscripttablewidth}{\dimexpr\linewidth-6\tabcolsep\relax}
\begin{tabular}{@{}L{0.225\manuscripttablewidth}C{0.310\manuscripttablewidth}C{0.180\manuscripttablewidth}C{0.285\manuscripttablewidth}@{}}
\toprule
\textbf{Well / Scope} & \textbf{Oil/Water Production RMSE (m\textsuperscript{3}/d) Baseline \ensuremath{\to} Round 5} & \textbf{Production Error Change} & \textbf{Water-Cut RMSE (pp) Baseline \ensuremath{\to} Round 5} \\
\midrule
\textbf{Field total} & 2,089.43 \ensuremath{\to} 544.72 & \ensuremath{-}73.9\% & --- \\
\textbf{P11} & 12.24 \ensuremath{\to} 3.75 & \ensuremath{-}69.4\% & 0.23 \ensuremath{\to} 0.07 \\
\textbf{P12} & 43.49 \ensuremath{\to} 9.48 & \ensuremath{-}78.2\% & 0.87 \ensuremath{\to} 0.20 \\
\textbf{P13} & 13.58 \ensuremath{\to} 5.81 & \ensuremath{-}57.2\% & 0.29 \ensuremath{\to} 0.12 \\
\textbf{P14} & 918.63 \ensuremath{\to} 277.92 & \ensuremath{-}69.7\% & 20.16 \ensuremath{\to} 6.11 \\
\textbf{P15} & 331.17 \ensuremath{\to} 150.53 & \ensuremath{-}54.5\% & 6.36 \ensuremath{\to} 2.79 \\
\textbf{P16} & 1,046.45 \ensuremath{\to} 446.06 & \ensuremath{-}57.4\% & 23.98 \ensuremath{\to} 10.19 \\
\bottomrule
\end{tabular}
\par\smallskip
\begin{minipage}{\linewidth}
\footnotesize\raggedright \textit{Note:} Combined field-and-well NRMSE fell by 69.4\% (37.98\% to 11.62\%), and active-sample well water-cut RMSE dropped from 12.90 to 4.91 pp. At 1\% water cut, P14--P16 broke through on days 1,101, 1,271, and 1,465 in Round 5 versus observed days 1,101, 1,240, and 1,358 (baseline delayed to days 2,126, 1,674, and 1,747).
\end{minipage}
\end{table}

\clearpage

\begin{figure}[H]
\centering
\includegraphics[width=\linewidth,height=428pt,keepaspectratio]{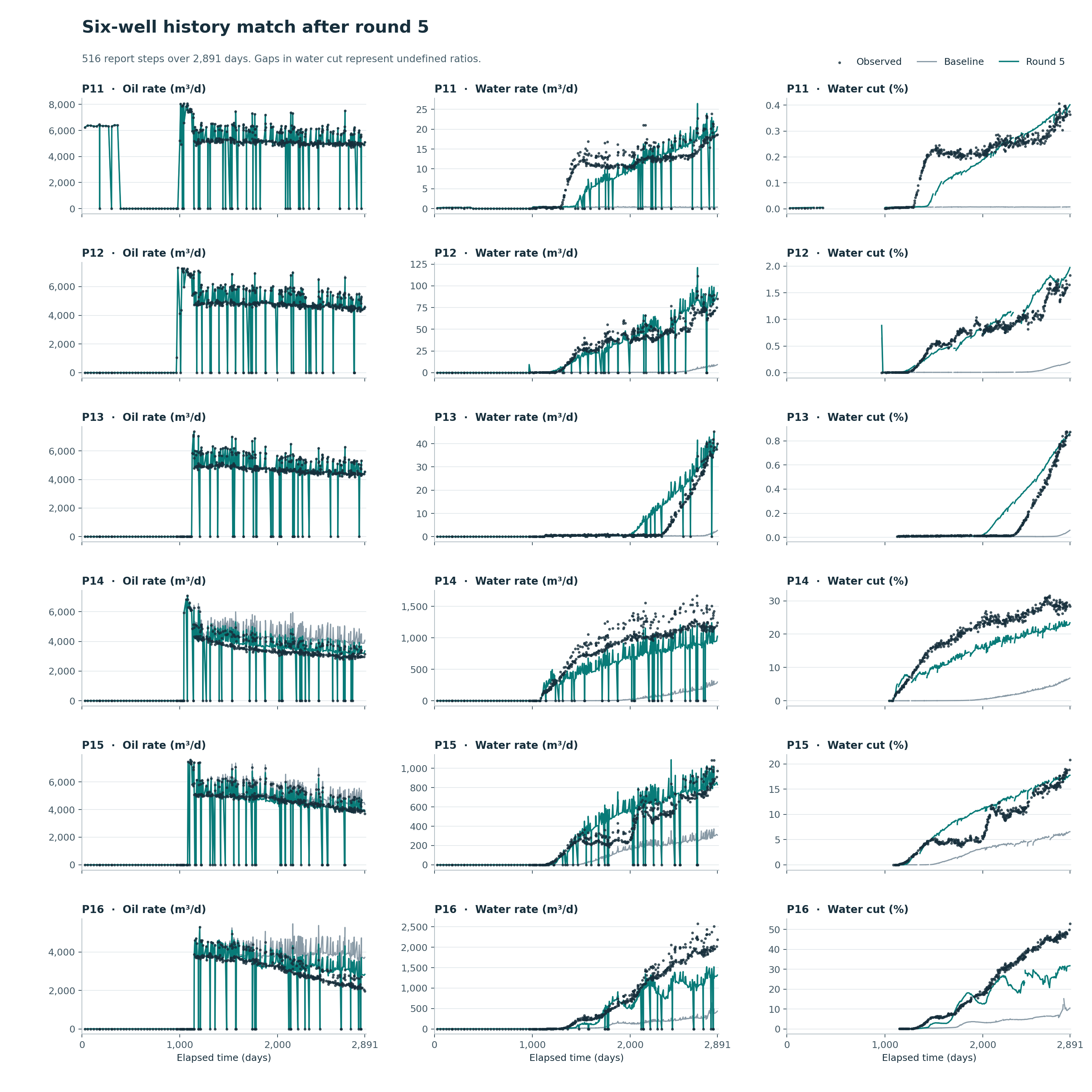}
\caption{Individual six-producer (P11--P16) oil production, water production, and water-cut history
match after Round 5 across 2,891 days in UNISIM-IV (Exp 04).}
\label{fig:unisim-well-profiles}
\end{figure}

By Round 5, low-water-cut wells (P11--P13) matched field history tightly (\textless{} 10
m\textsuperscript{3}/d production error and \ensuremath{\leq} 0.2 pp water-cut error), while
high-water-cut producers (P14--P16) achieved 55--70\% lower production errors with accurate
water-breakthrough timing (Fig.~\ref{fig:unisim-well-profiles} and Table~\ref{tab:unisim-errors}).
Independent auditing also confirmed that component mole fractions in every cell strictly satisfied the
unit-sum normalization constraint (Fig.~\ref{fig:unisim-spatial-edits}a). Remaining discrepancies
concentrate in late-time water production at P16 and P14, reflecting the five-round iteration limit and
non-uniqueness of full-field inversion under fixed geological layering.

\clearpage

\begingroup\ClosingSectionSpacing

\section{Discussion: Mechanistic Insights\texorpdfstring{\\}{ }and Paradigm Shifts in Reservoir Engineering}

In this study, we demonstrate the emerging ability of frontier large language models (LLMs) to natively
read and edit dynamic reservoir simulation models, establishing ``Subsurface Model Programming.'' Across
our experiments, the reasoning trajectories show that LLMs do not treat discrete reservoir grid cells
and model entries as isolated numerical symbols; instead, they construct coherent 3D physical
intuition---distinguishing sealed barriers from continuous channels, balancing reservoir pressure
support against channeling risk, and isolating flow conductivity from pore-volume sweep dynamics.
Equally important, providing clear physical feedback---such as well water-breakthrough timing and fluid
production evolution---rather than aggregate error metrics alone enables far more targeted adjustments
across high-dimensional parameter spaces. Although practical bottlenecks remain in context capacity for
massive reservoir files and in the computational runtime of iterative numerical simulations, these
capabilities signal a profound shift in subsurface interpretability and numerical optimization, pointing
toward four major future applications below.

\begin{figure}[H]
\centering
\includegraphics[width=\linewidth,height=403pt,keepaspectratio]{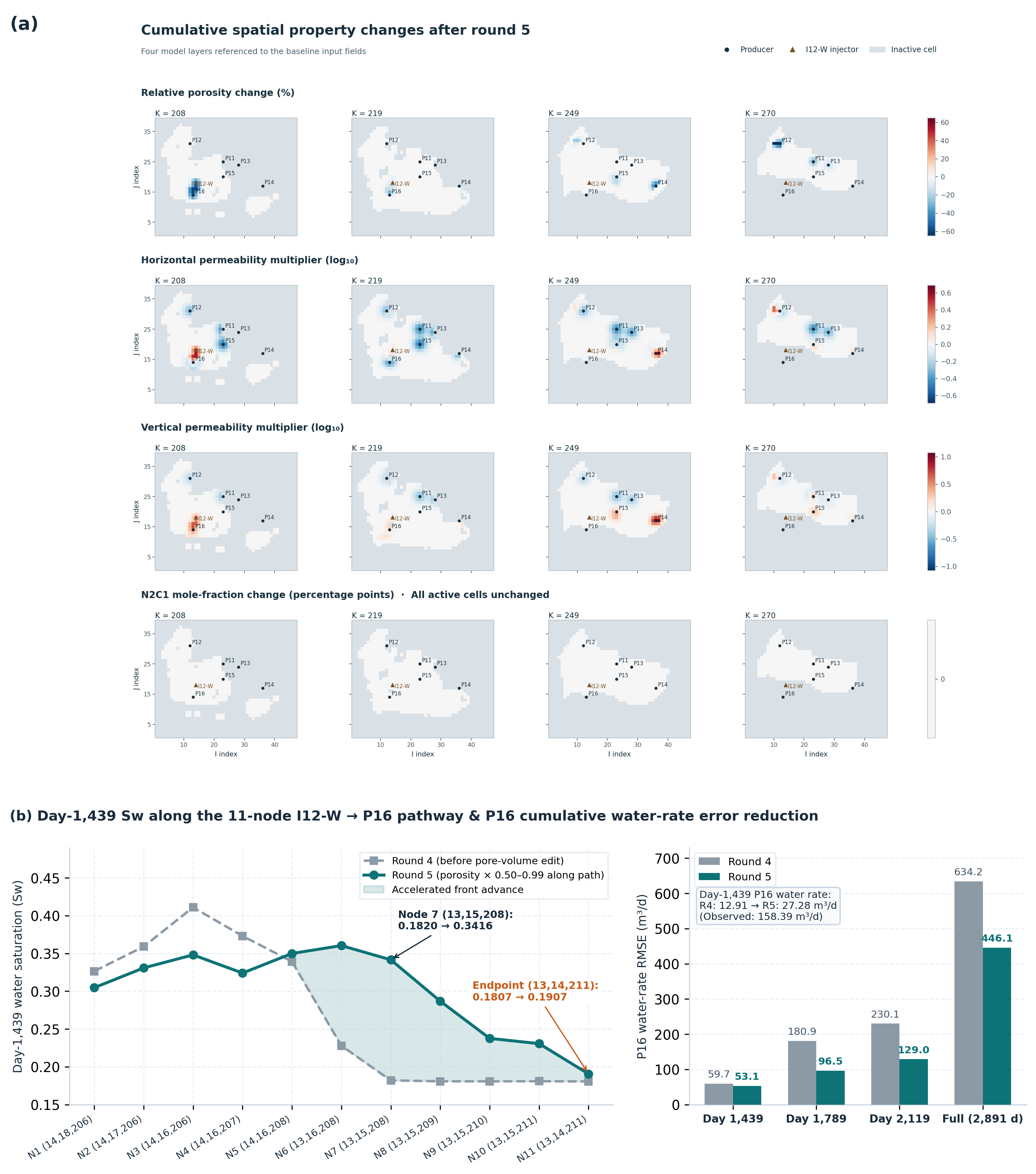}
\caption{Spatial localization and mechanistic verification of multi-round field edits in Exp 04. (a)
Cumulative changes in porosity, horizontal permeability, vertical permeability, and normalized initial
light-component mole fraction across four representative layers relative to baseline. (b) Day-1,439
water saturation advance along the 11-node discrete I12-W to P16 pathway triggered solely by Round 5's
porosity reduction, alongside progressive P16 water-production RMSE reduction across diagnostic
windows.}
\label{fig:unisim-spatial-edits}
\end{figure}

\subsection{Generative Synthesis and Physical Co-Optimization of Development Plans}

Conventional hydrocarbon development planning---spanning well patterns, well types, flooding strategies,
and staged controls---remains constrained by high forward-simulation costs and manual
pre/post-processing barriers. In routine engineering practice, teams often rely on localized experience
to propose only a small handful of candidate layouts followed by passive simulation checks, leaving vast
design spaces underexplored and struggling to anticipate complex inter-well interference in strongly
heterogeneous reservoirs \citep{Onwunalu2010,Humphries2014}. By natively parsing 3D structural highs,
permeability architecture, and remaining-oil saturation, LLMs internalize core engineering priors---such
as aligning injector--producer patterns with principal flow paths and controlling drawdown pressure to
suppress premature channeling---to synthesize physically sound development candidates efficiently.
Coupled with high-speed reservoir numerical simulators, this capability establishes a closed ``cognitive
synthesis -- rapid simulation -- physical feedback'' optimization engine, elevating field development
planning from empirical trial comparisons to physics-guided generative exploration.

\subsection{Physics-Guided Few-Shot History Matching and Geologically Plausible Inversion}

Mainstream ensemble and evolutionary inversion methods---such as Ensemble Kalman Filtering (EnKF),
smoother variants, and heuristic swarm search---typically demand \(O(10^{2})\)--\(O(10^{3})\) forward
simulations and apply diffuse parameter adjustments that often generate non-physical ``overfitting
noise'' distorting geological continuity
\citep{Onwunalu2010,Humphries2014,Naevdal2005,Oliver2011,Floris2001}. Native model programming turns the
LLM into a causal, prior-equipped inversion engine: attributing abnormal water-breakthrough timing and
water-production discrepancies directly to specific high-permeability thief channels or bottom-water
coning zones (Exp 02 and Exp 04). This mechanistic diagnosis compresses calibration simulation budgets
from \(O(10^{3})\) toward the \(O(10^{1})\) regime while preserving the geological continuity and
physical plausibility of the updated reservoir model.

\subsection{Automated White-Box Auditing and Knowledge Recovery for Legacy Reservoir Decks}

The petroleum industry maintains vast archives of legacy reservoir simulation input files accumulated
over decades of sequential multi-team edits. Many of these decks contain undocumented ``historical
patches''---such as arbitrary local transmissibility multipliers forced to fit past pressures,
unphysical vertical permeability overrides, or conflicting well-control switches---that silently
compromise forecast reliability. Capable of natively reading underlying simulation decks, LLMs can serve
as automated white-box auditors, scanning raw files directly to flag hidden physical violations,
diagnose numerical convergence stalls, recover undocumented modification intent, and generate clean,
conservation-verified remediation patches.

\subsection{Universal Native Simulation Interface for CCUS, Geothermal,\texorpdfstring{\\}{ }and Multiphysics Energy Systems}

Across emerging energy-transition workflows---including long-term CO\ensuremath{_{2}} plume tracking and
caprock integrity auditing in CCUS \citep{Class2009}, thermal breakthrough management in deep geothermal
reservoirs \citep{Pruess2004}, and underground hydrogen storage---modeling routinely requires coupling
multiphase flow, geochemistry, thermal transport, and geomechanics across black-oil, compositional,
thermal, and reactive-transport simulators \citep{Class2009,Pruess2004,Rasmussen2021,Moyner2025} with
disparate, rigid input syntaxes. Because Subsurface Model Programming operates directly on underlying
physical input files without relying on brittle, software-specific API wrappers, frontier LLMs establish
a universal syntax-agnostic interface for seamless cross-physics workflow orchestration and parameter
assimilation across the energy transition.

\section{Conclusions}

We demonstrate that frontier large language models (LLMs) can natively read and edit dynamic reservoir
simulation models, establishing ``Subsurface Model Programming'' as a physics-grounded computational
paradigm. Evaluated directly against open-source multiphase simulators across four progressively complex
benchmarks, LLM reasoning achieved 100\% accuracy in zero-tool 3D connectivity interpretation (Exp 01),
reduced PUNQ-S3 well-balanced water-cut error by 71.9\% in a single zero-tool pass editing just 3.1\% of
active cells (Exp 02), outperformed 25-evaluation PSO (+8.9\%) within five simulations on SPE10
development planning (Exp 03), and cut combined field-and-well normalized error by 69.4\% over five
rounds on the 72,267-active-cell UNISIM-IV compositional model (Exp 04).

By reasoning directly over underlying model physics rather than relying on blind trial-and-error,
Subsurface Model Programming bridges AI reasoning with physics-based reservoir simulation. Although
scaling long-context capacity, simulator runtime efficiency, and joint geological uncertainty remain
important frontiers, this paradigm lays a versatile foundation for autonomous development planning,
few-shot history matching, white-box auditing of legacy simulation assets, and broader energy-transition
applications including CO\ensuremath{_{2}} geological storage (CCUS) and deep geothermal systems.

\par\endgroup

\clearpage

\renewcommand{\refname}{References and notes}
\bibliographystyle{unsrtnat}
\bibliography{references}

\end{document}